\documentclass[%
 reprint,
 amsmath,amssymb,
 aps,
]{revtex4-2}

\usepackage{graphicx}
\usepackage{dcolumn}
\usepackage{bm}
\usepackage{hyperref}
\hypersetup{
        colorlinks=true,
        linkcolor=blue,
        filecolor=blue,
        urlcolor=blue,
        citecolor=blue
    }

\begin{document}

\preprint{APS/123-QED}

\title{Reciprocity of Charge-Orbital-Spin Transport in Normal-Metal/Ferromagnet Heterostructures}

\author{Abhishek Erram}
\thanks{Corresponding author Email: erramabhishek@gmail.com}%
\author{Akanksha Chouhan}
\author{Ashwin A. Tulapurkar}
\thanks{Corresponding author Email: ashwin@ee.iitb.ac.in}

\affiliation{Department of Electrical Engineering, Indian Institute of Technology Bombay, Mumbai 400076, India}

\date{\today}

\begin{abstract}
Orbital angular momentum has recently emerged as an important carrier of angular momentum in solids, offering new pathways for spin–orbitronic functionality beyond conventional spin transport. Here, we investigate the orbital Hall effect which generates orbital torques  and 
 their reciprocal process viz orbital pumping and the inverse orbital Hall effect (iOHE) in non-magnet/ferromagnet hetero-structures. Using two-port scattering-parameter measurements on Ru/Ni, Ru/Pt/CoFeB and Co/Cu/SiO$_2$ devices, we directly probe both orbital-torque-driven magnetization dynamics and orbital pumping within the same device platform. We observe that the transmission coefficients satisfy the symmetry relations required by Onsager reciprocity, demonstrating reciprocal conversion between charge, orbital and spin angular momenta. Our results establish orbital pumping as the reciprocal counterpart of orbital torque. Our experimental findings provide a unified framework for orbital transport phenomena.
\end{abstract}

\maketitle


Spin and orbital angular momenta are two important degrees of freedom of electrons and are fundamentally linked through spin–orbit coupling (SOC). The efficient generation of spin current is a central goal of spintronics because of its importance for magnetic memory, sensors, and related technologies \cite{1_fukami2025challenges,2_bhatti2017spintronics}. Materials with strong spin–orbit interaction (SOI), such as heavy metals as well as topological materials and transition-metal dichalcogenides (TMDs) \cite{3_liu2011spin,4_pai2012spin,5_liu2012spin,6_yan2017topological,7_fan2016electric,8_qi2011topological,9_hasan2010colloquium}, have therefore emerged as efficient platforms for charge-to-spin conversion. In these non-magnetic materials (NMs), spin-current generation is commonly attributed to the spin Hall effect (SHE) \cite{10_sinova2015spin,11_sanchez2013spin,12_manchon2019current} and related interfacial mechanisms \cite{13_manchon2015new,14_amin2020interfacial}. In spintronic systems, SOC plays a crucial role by enabling the interconversion between charge and spin degrees of freedom. Prominent examples include the spin Hall effect (SHE) and the Rashba–Edelstein effect (REE) \cite{10_sinova2015spin,15_hirsch1999spin,16_hirohata2020review} where an applied charge current gives rise to spin currents or spin accumulation. These effects form the physical basis of many spintronic technologies and are particularly important for nonvolatile magnetic random-access memory (MRAM), where information is written and read through spin-dependent transport.

Within the conventional spin–orbit framework, light metals with weak SOI have long been considered ineffective at generating sizable spin–orbit torques. Recent theoretical studies, however, predict that such materials can host strong orbital currents that do not rely on SOI \cite{17_jo2018gigantic,18_go2020theory,19_tanaka2008intrinsic,20_salemi2022first,21_go2018intrinsic}. These orbital currents constitute an independent and fundamental transport channel and can be converted into spin currents through spin–orbit coupling \cite{22_go2020orbital,23_lee2021orbital}. When injected into an adjacent ferromagnet, they can exert torques on the magnetization known as orbital torques, whose efficiency is determined by the electronic structure of the ferromagnetic layer. Experimental evidence for orbital-torque-driven phenomena has recently emerged \cite{24_bose2023detection,25_go2018intrinsic,26_dutta2022observation,27_lee2021efficient}, demonstrating that orbital transport can substantially enhance torque efficiency and broaden the material platform for spintronic devices \cite{28_shao2021roadmap,29_go2021orbitronics}. Unlike spin currents, orbital currents are not  constrained by relativistic effects, allowing them to reach magnitudes significantly larger than those of spin currents in a wide range of materials.

\begin{figure}
	\includegraphics[width=3.4in]{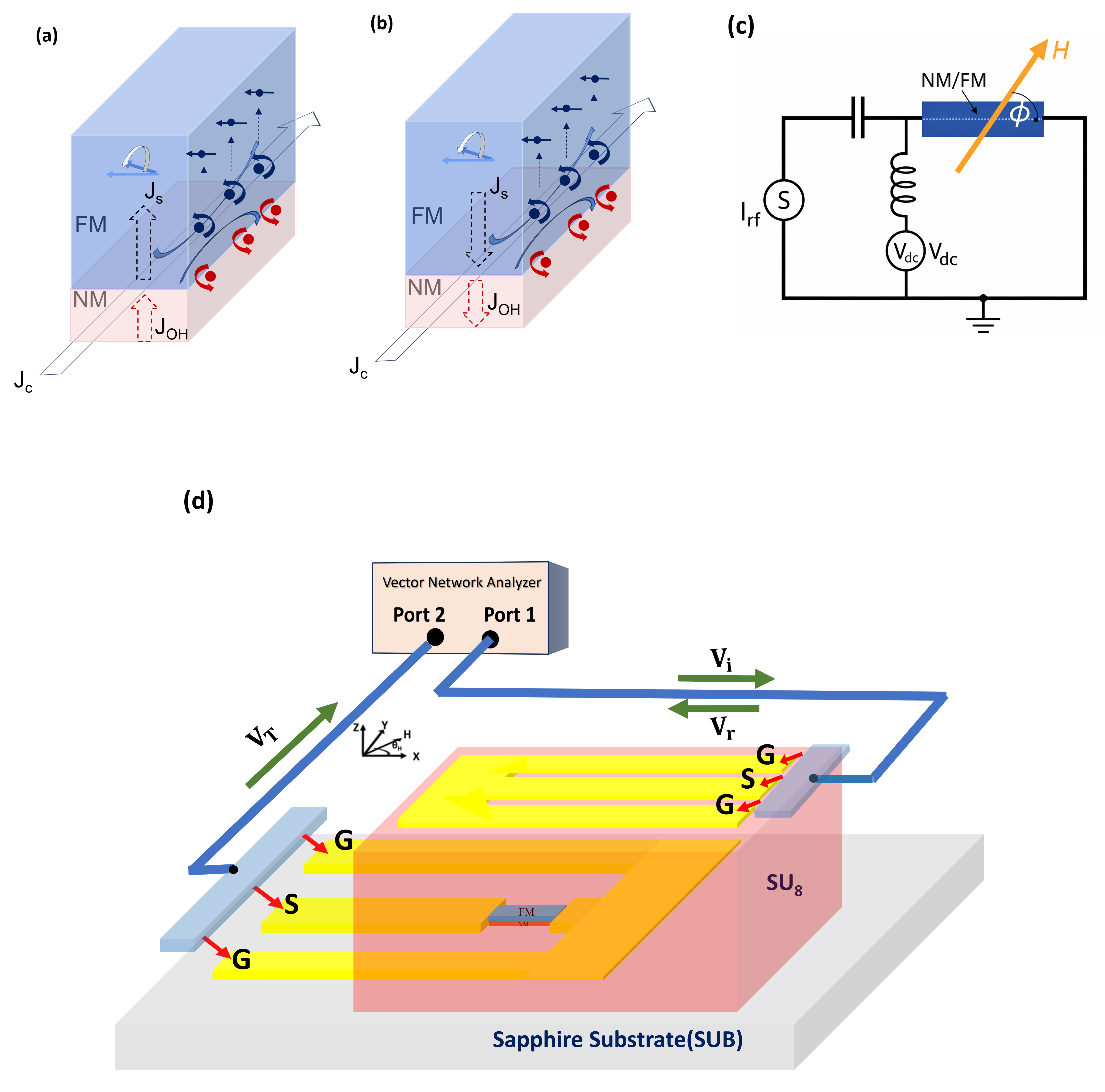}
    \caption{ Schematic illustration (a) Orbital current injection: an orbital current generated in the nonmagnetic layer (NM) is injected into the ferromagnet (FM) (blue arrow). At the NM/FM interface and within the FM, spin–orbit interaction (SOI) converts the injected orbital angular momentum into spin angular momentum (black arrow), which subsequently exerts an orbital torque on the ferromagnet, (b) Orbital pumping: a precessing ferromagnet injects angular momentum into an adjacent nonmagnetic metal (NM), where the spin current generated by magnetization dynamics is converted into an orbital current through the inverse orbital Hall effect (iOHE), (c) experimental setup of ST-FMR measurement, (d) Schematic representation of a two-port device used for S-matrix measurements. The bottom contacts of the NM/FM structure are connected to port 2, while port 1 consists of a coplanar waveguide that is electrically isolated from the bottom contacts. The S-matrix of the device is measured as a function of the applied external magnetic field. $V_i$, $V_r$, and $V_t$ denote the phasor amplitudes of the voltage waves incident on port 1, reflected from port 1, and transmitted to port 2, respectively. The scattering parameters $S_{11}$and $S_{21}$are defined as $S_{11}=V_{r}/V_{i}$ and $S_{21}=V_{t}/V_{i}$.}
	\label{fig1}
	
\end{figure}

When an orbital current generated by the orbital Hall effect (OHE) is injected into a ferromagnet, it can be converted into a spin current via the spin–orbit coupling within the ferromagnetic layer, thereby exerting an orbital torque on the magnetization. This is shown schematically in Fig. \ref{fig1}a. Because this orbital-to-spin conversion primarily occurs inside the ferromagnet rather than in the NM, the process constitutes nonlocal orbital transport \cite{25_go2018intrinsic,26_dutta2022observation}. In some prototype NM/HM/FM devices, the spin–orbit coupling of the HM is used to convert the orbital current to a spin current, further enhancing spin-torque efficiency \cite{24_bose2023detection,30_kim2021nontrivial,31_ding2020harnessing,32_ding2022observation,42_kim2023oxide}. Apart from  generating orbital currents through the OHE,  the orbital Rashba–Edelstein effect (OREE) \cite{1_fukami2025challenges} where a non-equilibrium orbital angular momentum is accumulated  at the interface between two materials, has been utilized for generating orbital torques.

Orbital pumping (fig.\ref{fig1}b) which is reciprocal to the orbital Hall torque has been proposed \cite{46_go2025orbital} and tested experimentally, in which the dynamic precession of a magnet emits a pure orbital current without an accompanying charge current \cite{33_keller2025identification,34_hayashi2024observation}. However these experimental studies have not explored the Onsager reciprocity between orbital Hall torque and orbital pumping.   Recently, some studies have examined the  reciprocity between charge-to-orbital and orbital-to-charge   conversion within the same device\cite{41_gao2025nonlocal},\cite{40_mendoza2024efficient}, \cite{35_ledesma2025nonreciprocity}. Reciprocity between OREE and its inverse was demonstrated in ref \cite{41_gao2025nonlocal} by using non-local electrical detection method. However, the work in ref \cite{40_mendoza2024efficient}and \cite{35_ledesma2025nonreciprocity}, which is based on magnon mediated transport, found that the efficiencies of charge-to-orbital and orbital-to-charge interconversion are different, implying non-reciprocity. A recent study on W/Ni bilayer system has shown violation of local reciprocity by comparing results from STFMR and spin-pumping experiments \cite{47_kashiki2026violation}. 

We here demonstrate that the effects based on OHE (or OREE) and orbital torque and the inverse processes viz. i-OHE (or i-OREE) and orbital pumping  follow Onsager reciprocity. We measure the scattering parameters of two-port devices where the transmission of electrical signal from port 2 to 1 is based on orbital torques generated via OHE (or OREE), whereas the transmission from port 1 to 2 is based on orbital-pumping and i-OHE (or i-OREE) effects (fig.\ref{fig1}d). 
The measurements are done in rf frequency range, where the transmission is enhanced when ferromagnetic resonance is excited.
   
 \begin{figure}
	\includegraphics[width=3.4in]{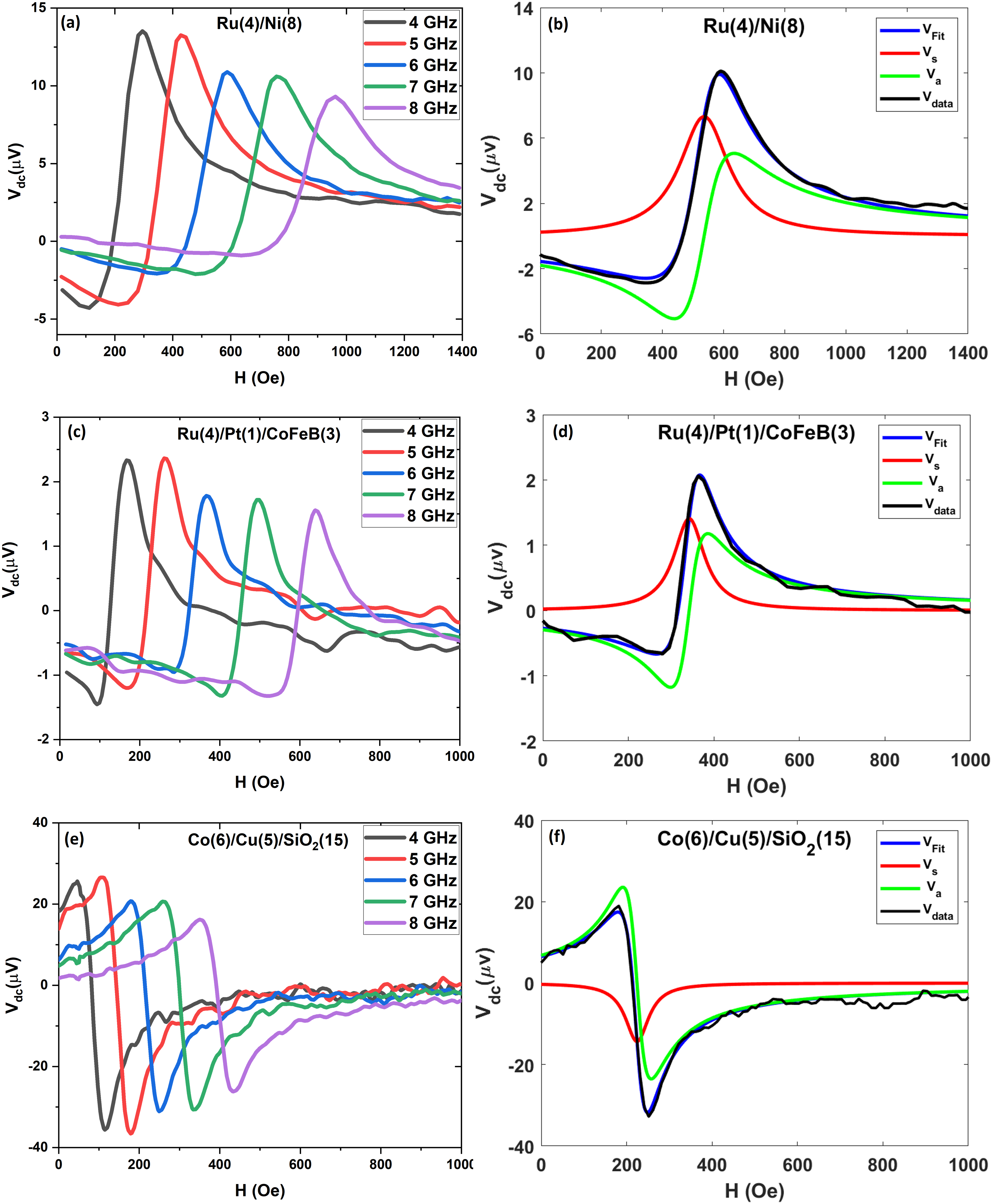}
 
	\caption{(a),(c),(e) ST-FMR voltage signals at various frequencies as a function of an in-plane external magnetic field applied at $45^o$ angle for Ru(4)/Ni(8), Ru(4)/Pt(1)/CoFeB(3) and Co/Cu/SiO$_2$ respectively. (b),(d),(f) The data at 6GHz are fitted with a combination of symmetric Lorentzian functions $V_{S}$ (Red curve) and antisymmetric Lorentzian functions $V_{A}$ (Green curve) }
	\label{fig2}
\end{figure}    

In this work we have prepared three sets of samples: (a) sapphire/ Ta(1)/ Ru(4)/ Ni(8)/ MgO(1.5)/ Ta(1.2), (b) sapphire/ Ta(1)/ Ru(4)/ Pt(1)/ CoFeB(3)/ MgO(1.5)/ Ta(1.2), and (c) sapphire/ Co(6)/ Cu(5)/ SiO$_2$(15), where the numbers in parentheses denote thicknesses in nanometers (nm). The MgO(1.5)/ Ta(1.2) bilayer served as a capping layer to prevent oxidation of the ferromagnetic layer in first two samples. Sapphire substrates were used to minimize RF losses. All films were deposited by magnetron sputtering at a base pressure below $1 \times 10^{-8}$~Torr. 
Rectangular strips with typical dimensions of $10 \times 100~\mu\mathrm{m}^2$ were defined by optical lithography, followed by sputter deposition and lift-off. Bottom coplanar waveguide structures were then fabricated, followed by sputter deposition of Ti/Au. Subsequently, an SU-8 (2000.5) dielectric layer was patterned and hard-baked at 150~$^\circ$C for 25~min, resulting in a thickness of approximately 3--4~$\mu$m. Finally, the top waveguide and ground--signal--ground (GSG) contact pads were fabricated, followed by sputter deposition of Ti/Au, as shown in Fig. \ref{fig1}(d).

First we performed spin torque ferromagnetic resonance (ST-FMR) measurements on  Ru/Ni,Ru/Pt/CoFeB and Co/Cu/SiO$_2$\cite{24_bose2023detection,3_liu2011spin,36_pai2015dependence,37_tulapurkar2005spin} devices using the port 2,  to verify that the ferromagnetic layers can be excited by angular-momentum currents.  
A radio-frequency (rf) current with frequency $f_o$= 4-8 GHz at 10 dBm power was applied to the device while sweeping an in-plane external magnetic field oriented at 45° with respect to the strip axis as shown in Fig. \ref{fig1}(c). 
The rf current generates an Oersted field ($H_{oe}$) and orbital currents due to the the orbital Hall effect of  Ru (sample a and b). The orbital current injected into Ni (sample a) is converted into spin current due to the spin-orbit interaction of Ni and exerts a spin torque on Ni.  In the case of sample b, the orbital current is injected into Pt layer and converted into spin-current due to the spin-orbit interaction of Pt. This spin current is then injected into the CoFeB layer and exerts a spin torque.   In the case of sample c, the orbital Rashba-Edelstein effect at the Cu-SiO$_2$ interface generates non-equilibrium orbital angular momentum accumulation. This drives an orbital current, which is converted into spin current by Co and results in spin-torque. Thus in all the samples, rf current drives the magnetization oscillation through Oersted field and orbital torques. 
The homodyne mixing of the oscillating resistance arising from anisotropic magnetoresistance (AMR) with the rf current produces a dc voltage signal. This signal after subtraction of a constant background, can be described by a superposition of symmetric ($V_S$) and antisymmetric ($V_A$) Lorentzian line shapes:
\begin{equation}
V_S(H)
=
C_1 \,
\frac{\Delta^2}
{4(H - H_r)^2 + \Delta^2}
\end{equation}

\begin{equation}
V_A(H)
=
C_2 \,
\frac{2\Delta (H - H_r)}
{4(H - H_r)^2 + \Delta^2}
\end{equation}

\begin{figure}
	\includegraphics[width=3.4in]{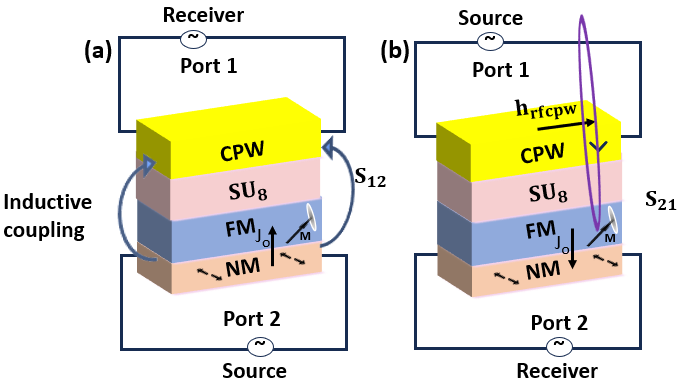}
	\caption{The device schematic used to show the reciprocity between (a) OHE $+$ Orbital Torque and (b) Orbital pumping $+$ iOHE}
	\label{fig3}
\end{figure}

where $\Delta$ is the linewidth, $H_r$ is the resonance field, $C_1$  is the strength of damping-like torque and $C_2$ is the strength of field-like torque. In all the samples, 
$V_S$ arises from in-plane damping-like torque excitation generated by orbital currents 
while $V_A$ originates from out-of-plane field-like torques arising from the Oersted field of the rf current.
Fig. \ref{fig2}a, \ref{fig2}c, \ref{fig2}e shows the ST-FMR plots in the 4-8 GHz frequency range for samples a, b and c respectively.
At 6 GHz, we obtained $\Delta =200.7\; Oe, H_{r}=572.6\; Oe$
for Ru/Ni system, $\Delta = 87.15\; Oe, H_{r}=338\; Oe$
for Ru/Pt/CoFeB system and $\Delta =66.44\; Oe, H_{r}=224.3\; Oe$
for Co/Cu/SiO$_2$ system, as show in fig.\ref{fig2}(b), \ref{fig2}(d) and \ref{fig2}(f). These ST-FMR measurements confirm efficient current-induced excitation of ferromagnets in all the three hetero-structures.

\begin{figure}
	\includegraphics[width=3.4in]{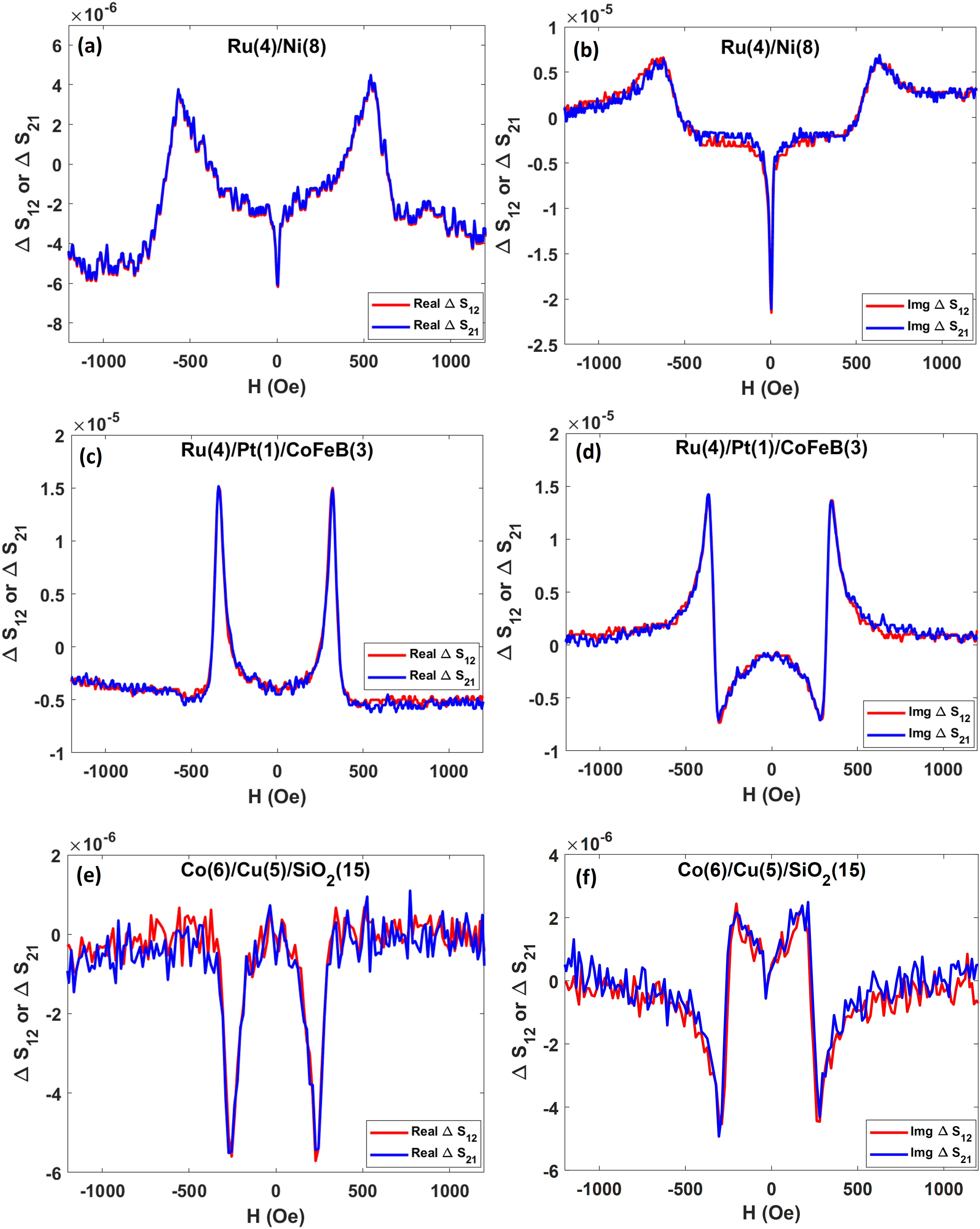}
    
	\caption{The real and imaginary parts of the transmission coefficients $\Delta$$S_{12}$ and $\Delta$$S_{21}$ are plotted as a function of the applied external magnetic field after background subtraction. Both $\Delta S_{12}$ and $\Delta S_{21}$ are symmetric with respect to the magnetic field for the Ru/Ni sample (a), (b) , the Ru/Pt/CoFeB sample (c), (d) and Co/Cu/SiO$_2$ (e), (f).}
	\label{fig4}
\end{figure}
Next, we measured the scattering matrix (S-matrix) of the devices using a vector network analyzer (VNA) \cite{44_shukla2020generation, 45_bhuktare2017gyrator}. The top coplanar waveguide is taken as port 1 and the bottom waveguide connected to the FM/NM multilayer is taken as port 2, as shown in Fig. \ref{fig1}(d). The measurements were performed at 6 GHz frequency and input power of 10 dBm.
An in-plane magnetic field was swept along the length of the microstrip, and the S-parameters were recorded as a function of the magnetic field. These two-port measurements were used to investigate reciprocal conversion between charge, orbital, and spin angular momenta. 

Let's first consider the Ru/Ni device.  The $S_{12}$ parameter is the ratio of ac voltage detected at port 1 to the ac voltage applied to port 2. This transmission of signal from port 2 to port 1 takes place by the following steps: (1) When ac voltage is applied to port 2, ac charge current flows through the Ru/Ni multi-layer. The charge current flowing through Ru creates orbital current via OHE. (2) The orbital current is converted into spin current inside Ni due to the spin-orbit coupling and it excites the magnetization via spin-torque.  (3) The oscillating magnetization induces voltage in the top waveguide via the inductive coupling (Faraday's law of electromagnetic induction). These processes are shown schematically in Fig. \ref{fig3}a. The transmission of signal also takes place by a second route: The magnetization is excited by the Oersted magnetic field created by charge current (ampere's law) and step 3 above.

The $S_{21}$ parameter is the ratio of ac voltage detected at port 2 to the ac voltage applied to port 1. This transmission of signal from port 1 to port 2 takes place by the following steps: (1') When ac voltage is applied to port 1, ac charge current flows through the top wave guide. This creates ac magnetic field (Ampere's law) which excites the magnetization of Ni. (2') The oscillating magnetization pumps ac spin-current, which is converted into ac orbital current due to the spin-orbit coupling of Ni.  (3') The ac orbital current injected in the Ru layer is converted to charge current via the inverse OHE and detected as voltage signal at port 2.  These processes are shown schematically in Fig. \ref{fig3}b. Here also, the signal transmission takes place by a second route: step 1' above and oscillating magnetization induces ac voltage in the waveguide 2 via the Faraday's law.

The real and imaginary parts of $\Delta S_{12}$ and $\Delta S_{21}$ parameters as a function of magnetic field for Ru/Ni device are shown in Fig. \ref{fig4}(a,b). ( $\Delta S$ is obtained by subtracting value of S parameter at high field from data.) We can see that the signals are enhanced near the ferromagnetic resonance field. The signals are symmetric w.r.t. the magnetic field and therefore obey the generalized Onsager reciprocity viz $S_{21}(m,H)=S_{12} (-m,-H)$. This implies that the three steps involved in the transmission from port 2 to 1, and from port 1 to 2 are reciprocal of each other i.e. step 1 (OHE) and step 3'(i-OHE) are reciprocal, step 2 (orbital torque) and step 2'(orbital pumping) are reciprocal and step 3 (Faraday's law) and step 1' (Ampere's law) are reciprocal. The steps in the second routes of transmission of signals are reciprocal as well.

Now let's consider transmission of signal from port 2 to 1 in the Ru/Pt/CoFeB device. Here the orbital current is converted into spin current by the Pt layer instead of the ferromagnetic layer. All other steps for transmission are same as the previous case of Ru/Ni. In the transmission of signal from port 1 to 2, the difference from the previous case is that the ac spin current pumped by the FM is converted into orbital current by the Pt layer via spin-orbit coupling. The real and imaginary parts of $\Delta S_{12}$ and $\Delta S_{21}$ parameters as a function of magnetic field for Ru/Pt/CoFeB device are shown in Fig. \ref{fig4}(c,d).  In this case also, the signals are symmetric w.r.t. the magnetic field and  obey the generalized Onsager reciprocity viz. $S_{21}(m,H)=S_{12} (-m,-H)$.
Finally, let's consider transmission of signal from port 2 to 1 in the Co/Cu/SiO$_2$ device. Here the orbital current is generated via the OREE at the Cu/SiO$_2$ interface. The remaining steps are same as Ru/Ni device i.e. the orbital current is converted into spin current by FM etc. In the transmission of signal from port 1 to 2, the pumped orbital current is converted into charge current via the inverse OREE at  Cu/SiO$_2$ interface. The other steps are the same as Ni/Ru device.  The real and imaginary parts of $\Delta S_{12}$ and $\Delta S_{21}$ parameters as a function of magnetic field for  Co/Cu/SiO$_2$ device are shown in Fig. \ref{fig4}(e,f).  In this case also, the signals are symmetric w.r.t. the magnetic field and  obey the generalized Onsager reciprocity viz. $S_{21}(m,H)=S_{12} (-m,-H)$.

\begin{figure}
	\includegraphics[width=3.4in]{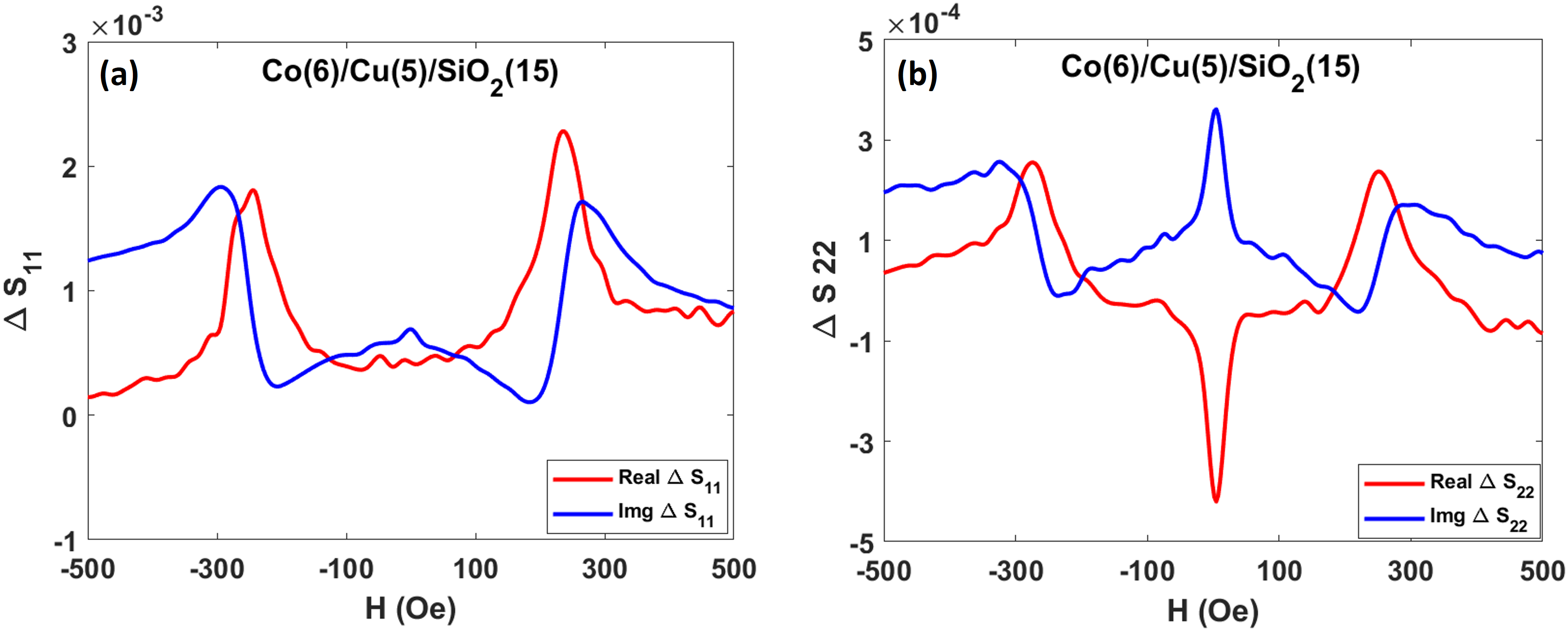}
	\caption{$\Delta S_{11}$ and $\Delta S_{22}$ real part shows a peak at resonance and imaginary part shows dispersion behavior.}
	\label{fig5}
\end{figure}

The real and imaginary parts $\Delta S_{11}$ and $\Delta S_{22}$ parameters (which are reflection coefficients) are plotted in Fig. \ref{fig5}(a,b) for device c. We can see that the real part shows a peak at resonance where as the imaginary part shows dispersion. This is expected as the real part corresponds to dissipation and therefore should be positive. The dissipation here arises from the oscillation of magnetization. Let's consider the $S_{22}$ parameter. This corresponds to the reflection of  wave incident on port 2. The reflected wave can be thought of as a wave generated by the device in response to the incident wave.
The incident wave oscillates the magnetization, which  gives rise to orbital pumping and finally a voltage wave traveling back is generated. Thus similar to $S_{12}$ parameter, the $S_{22}$ parameter also involves various steps and is a combination of OHE, orbital torques, orbital pumping and inverse-OHE. All these reciprocal processes together act in such a way that the reflected wave results in positive dissipation.
The shape of $S_{12}$ (and $S_{21}$) parameters are  a combination of peak and dispersion (fig.\ref{fig4}). It can be seen that the imaginary part of $\Delta S_{12}$ has a shape similar to the ST-FMR signals.
For all the three devices, by varying the input power from 0 to 10 dBm, we confirmed that all S-parameters remain independent of power, indicating operation within the linear-response regime \cite{38_bhuktare2019direct}.

We have demonstrated  the reciprocity between charge-to-orbital and orbital-to-charge conversion in three different type of devices comprising multi-layers of  Ru/Ni, Ru/Pt/CoFeB and Co/Cu/SiO$_2$. The orbital current plays an essential role in the devices. We have shown that the scattering parameters in these devices follow Onsager reciprocity relations. Our results show that orbital torque and orbital pumping are reciprocal processes.

\section*{Acknowledgement}
Authors are thankful for the support provided by
IIT Bombay Nanofabrication Facility, Indian Institute of
Technology Bombay. We also acknowledge the
support of The Ministry of Electronics and Information Technology,
Government of India through Project: Translation and Research for Advanced Nanoelectronics Growth (TaRANG).

\bibliography{refs}
\end{document}